\documentclass[showpacs,superscriptaddress,twocolumn,aps, longbibliography]{revtex4-2} 
\usepackage[USenglish]{babel}
\usepackage{graphicx,xcolor}
\usepackage{amsmath, bbm}
\usepackage{mathtools}
\usepackage{natbib}
\usepackage[hidelinks]{hyperref}
\usepackage{epstopdf}
\usepackage{color}
\usepackage{notes2bib}
\usepackage{siunitx}
\usepackage{relsize}

\newcommand{\Ostk}{\Omega_{\mathrm{stk}}}
\newcommand{\Oge}{\Omega_{ge}}
\newcommand{\Pstk}{P_{\mathrm{stk}}}
\newcommand{\Dstk}{\Delta_{\mathrm{stk}}}
\newcommand{\Oexc}{\Omega_{\mathrm{exc}}}
\newcommand{\Pexc}{P_{\mathrm{exc}}}
\newcommand{\Dexc}{\Delta_{\mathrm{exc}}}
\newcommand{\Dge}{\Delta_{ge}}

\newcommand{\lesssimsmall}{\,\mathsmaller{\lesssim}\,}

\begin{document}
\title{Spectral splitting of a stimulated Raman transition in a single molecule}
 \author{Johannes Zirkelbach}%
\affiliation{Max Planck Institute for the Science of Light, D-91058 Erlangen, Germany}
\affiliation{Friedrich-Alexander-Universität Erlangen-Nürnberg, Department of Physics, D-91058 Erlangen, Germany}
 \author{Burak Gurlek}%
\affiliation{Max Planck Institute for the Science of Light, D-91058 Erlangen, Germany}
\affiliation{Friedrich-Alexander-Universität Erlangen-Nürnberg, Department of Physics, D-91058 Erlangen, Germany}
 \author{Masoud Mirzaei}%
\affiliation{Max Planck Institute for the Science of Light, D-91058 Erlangen, Germany}
\affiliation{Friedrich-Alexander-Universität Erlangen-Nürnberg, Department of Physics, D-91058 Erlangen, Germany}
\author{Alexey Shkarin}%
\affiliation{Max Planck Institute for the Science of Light, D-91058 Erlangen, Germany}
\author{Tobias Utikal}%
\affiliation{Max Planck Institute for the Science of Light, D-91058 Erlangen, Germany}
\author{Stephan G\"otzinger}%
\affiliation{Friedrich-Alexander-Universität Erlangen-Nürnberg, Department of Physics, D-91058 Erlangen, Germany}
\affiliation{Max Planck Institute for the Science of Light, D-91058 Erlangen, Germany}
\affiliation{Graduate School in Advanced Optical Technologies (SAOT), Friedrich Alexander
University Erlangen-Nuremberg, D-91052 Erlangen, Germany}
\author{Vahid Sandoghdar}
\affiliation{Max Planck Institute for the Science of Light, D-91058 Erlangen, Germany}
\affiliation{Friedrich-Alexander-Universität Erlangen-Nürnberg, Department of Physics, D-91058 Erlangen, Germany}

\date{\today}

\begin{abstract}
The small cross section of Raman scattering poses a great challenge for its direct study at the single-molecule level. By exploiting the high Franck-Condon factor of a common-mode resonance, choosing a large vibrational frequency difference in electronic ground and excited states and operation at $T< 2\,K$, we succeed at driving a coherent stimulated Raman transition in individual molecules. We observe and model a spectral splitting that serves as a characteristic signature of the phenomenon at hand. Our study sets the ground for exploiting the intrinsic optomechanical degrees of freedom of molecules for applications in solid-state quantum optics and information processing.
\end{abstract} 

\maketitle

On February 28, 1928, Raman recorded spectra that provided conclusive evidence for inelastic coherent scattering in the optical domain \cite{raman-1928-b, raman-1928}. Today, we know that the Raman process is due to the coupling of the atomic vibrations in a molecule to its electronic transitions, which can be seen as the natural realization of a quantum optomechanical phenomenon \cite{aspelmeyer-2014, benz-2016, velez-2019}. The effect has found a wide-spread interest for its molecular fingerprinting capacity in sensing and diagnostics applications \cite{kudelski-2008, tu-2012}. 
These investigations typically involve ensembles of molecules since cross sections of Raman scattering are extremely small although single-molecule sensitivity has been reported in schemes based on near-field plasmonic enhancement \cite{nie-1997, kneipp-1997, le-2012, benz-2016, zrimsek-2017, zong-2019}.
In a recent study, Xiong et al.~achieved single-molecule Raman microscopy with chemical sensitivity at room temperature using a clever electronically pre-resonant stimulated scheme \cite{xiong-2019}. In this Letter, we demonstrate robust single-molecule Raman signals obtained at high spectral resolution in a coherent and fully resonant configuration under cryogenic conditions. 


In 1989, Moerner and Kador showed that single PAHs embedded in crystalline matrices could be addressed at liquid helium temperature via Fourier-limited zero-phonon lines (00ZPL) connecting the vibrational ground levels of the electronic ground and excited states \cite{moerner-1989}. Since then, a wide range of optical experiments have exploited 00ZPLs of PAHs for quantum optical and nonlinear optical investigations \cite{tamarat-1999, maser-2016, toninelli-2021, adhikari-2022-2}. However, the vibrational levels of the electronic states have only been studied via incoherent spectroscopy in these systems \cite{plakhotnik-2002, banasiewicz-2007, wiacek-2008, zirkelbach-2022}. In what follows, we show that despite short lifetimes of the order of picoseconds, it is possible to drive resonant stimulated Raman transitions to vibrational levels of the electronic ground state using continuous-wave (CW) laser radiation, leading to the observation of spectral splitting. Our work paves the way for coherent quantum optical operations using the vibrational degrees of freedom in molecules.
\begin{figure}[t]
\centering 
 \includegraphics[width=0.9\columnwidth]{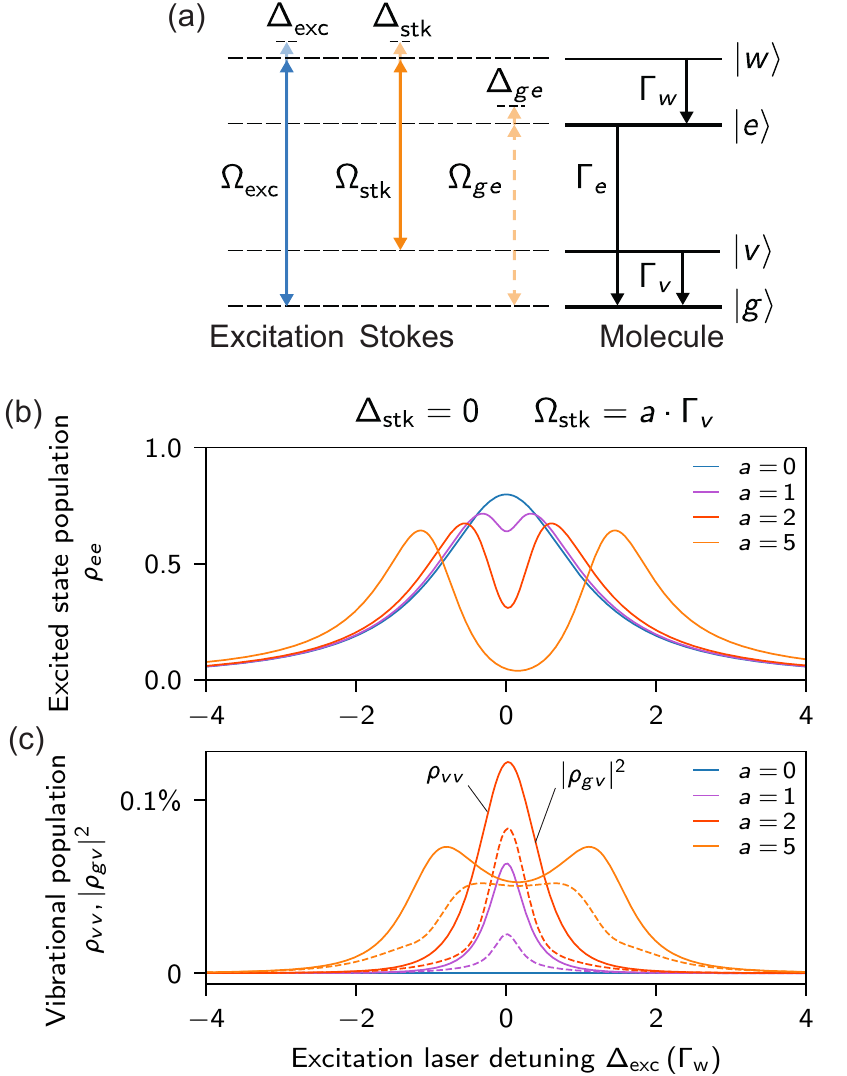}
 \caption{(a) Level scheme for model calculations including the electronic ground and excited states $\vert g \rangle$ and $\vert e \rangle$ with no vibrational excitation as well as the vibronic states $\vert v \rangle$ and $\vert w \rangle$. 
Arrows indicate which transitions can be reached with the excitation laser (blue) and Stokes laser (yellow).
 $\Gamma_{i}$: Relaxation rates, $\Omega_{i}$: Rabi frequencies, $\Delta_{i}$: frequency detunings.
 (b) Population in state $\vert e \rangle$ for
 $\Gamma_w = 2 \Gamma_v = 1000\Gamma_e$, 
 $\Oexc = 2\sqrt{\Gamma_{e}\Gamma_{w}}$, $(\nu_{vw} - \nu_{ge})/(2\pi) = -10\Gamma_{w}$, $\Oge=\Ostk$, $\Dstk = 0$, and various values of the Stokes laser Rabi frequency $\Ostk = a \cdot \Gamma_v$. The values are plotted as a function of the excitation laser detuning $\Dexc$. 
 For $\Ostk = 5 \Gamma_v$, the profile is not centered around $\Dexc=0$ because the Stokes laser exerts a noticeable AC-stark shift on the $\vert g \rangle \leftrightarrow \vert e \rangle$ transition. (c) Same as in (b) but for $\rho_{vv}$ (solid curves) and its coherent part $\vert \rho_{gv} \vert^2$ (dashed curves).}
 \label{model} 
\end{figure}

We begin with a brief theoretical discussion of the interaction between two laser fields and a generic four-level molecule \cite{shi-2018, wei-2018, xiong-2019, saikan-1985, pei-2013}, in line with other treatments in the broader context of coherent optical processes such as electromagnetically induced transparency (EIT) and Autler-Townes splitting (ATS) \cite{fleischhauer-2005, lazoudis-2010, ahmed-2012}. Figure\,\ref{model}(a) depicts the general level scheme, where $\vert g \rangle$ and $\vert e \rangle$ denote the molecular electronic ground and excited states, respectively, with no vibrational excitation. States $\vert v \rangle$ and $\vert w \rangle$ stand for their vibrational levels with much shorter lifetimes due to rapid phonon-assisted relaxation in the solid matrix \cite{hill-1988, hill-1988-5, califano-1981}. Parameters $\Gamma_{i}$ represent the decay rate of state $\vert i \rangle$, whereby in our system $\Gamma_{w} \approx  2\Gamma_{v}\gg\Gamma_{e}$. 

In the rotating-wave approximation, the Hamiltonian of the model system can be written as \cite{lazoudis-2010}  
\begin{equation} \label{hamiltonian}
    \hat{\tilde{H}} =
    \hbar
    \begin{pmatrix}
    0 & 0 &  \Omega_{ge}/2 & \Omega_{\mathrm{exc}}^{}/2 \\
    0 & \Delta_{\rm stk}- \Delta_{\rm exc} & 0 & \Omega_{\mathrm{stk}}^{}/2 \\
    \Omega_{ge}/2 & 0 & -\Delta_{ge} & 0 \\
    \Omega_{\mathrm{exc}}^{}/2 & \Omega_{\mathrm{stk}}^{}/2 & 0 & -\Delta_{\mathrm{exc}}
    \end{pmatrix}
\end{equation}
in the rotating frame with the basis $\{\vert g \rangle, \vert v \rangle, \vert e \rangle, \vert w \rangle \}$. Here, $\Omega_{\rm exc} = - \vec{d}_{gw} \cdot \vec{E}_{\rm exc}/\hbar $ and $\Omega_{\rm stk} = - \vec{d}_{vw} \cdot \vec{E}_{\rm stk}/\hbar$ denote the Rabi frequencies produced by the excitation (exc) and Stokes (stk) laser fields, $\vec{E}_{\rm exc}$ and $\vec{E}_{\rm stk}$, respectively, and $\vec{d}_{ij}$ is the transition dipole moment between states $\vert i \rangle$ and $\vert j \rangle$. In addition, $\Oge = \beta_g \Ostk$, where $\beta_g=\lVert \vec{d}_{ge} \rVert / \lVert \vec{d}_{vw} \rVert$ represents the relative Franck-Condon (FC) overlap between the 00ZPL and the $\vert v \rangle \leftrightarrow \vert w \rangle$ transition. We have included the $\vert g \rangle \leftrightarrow \vert e \rangle$ transition in the model because of its spectral vicinity to $\vert w \rangle \leftrightarrow \vert v \rangle$ transition addressed by the Stokes laser. The parameters $\Dexc = 2\pi (\nu_{\mathrm{exc}} - \nu_{gw})$, $\Dstk=2\pi (\nu_{\mathrm{stk}} - \nu_{vw})$ and $\Dge = 2\pi (\nu_{\mathrm{stk}} - \nu_{ge})$ signify various frequency detunings. Here, $\nu_{ij}$ stands for the frequency of the $\vert i \rangle \leftrightarrow \vert j \rangle$ transition, and $\nu_{\rm exc}$ and $\nu_{\rm stk}$ are the frequencies of the excitation and Stokes lasers, respectively (see Supplemental Material (SM), sections II and III). 
From a quantum optical perspective, such a system is situated between the EIT and ATS regimes for moderate driving of the $\vert v \rangle \leftrightarrow \vert w \rangle$ transition ($\Ostk \lesssim \Gamma_w$) \cite{anisimov-2011, lu-2015, hao-2018}.

Considering that $\vert w \rangle$ decays very fast, the population $\rho_{ee}$ of $\vert e \rangle$ reports on the excitation of the molecule via the $\vert g \rangle \leftrightarrow \vert w \rangle$ transition.  In Fig.\,\ref{model}(b), we portray the numerical solution of the Lindblad master equation for $\rho_{ee}$. 
These spectral profiles result from a line splitting and a dip that forms in the absorption profile of the transition $\vert g \rangle \leftrightarrow \vert w \rangle$ 
as coherent population exchange via the vibronic $\vert w \rangle \leftrightarrow \vert v \rangle$ transition becomes comparable to the relaxation rate of $\vert v \rangle$ and $\vert w \rangle$, i.e., as $\Ostk$ becomes comparable to $\Gamma_{v}, \Gamma_{w}$.
Indeed, $\rho_{ww}$ follows the same spectral profile as $\rho_{ee}$, albeit with a much smaller amplitude. The quantities $\rho_{vv}$ (solid curves) and $\vert \rho_{gv} \vert^2$ (dashed curves) in Fig.\,\ref{model}(c) show that a small fraction of the population is coherently transferred to $\vert v \rangle$, establishing a stimulated Raman transition from $\vert g \rangle$ to $\vert v \rangle$. The correspondence between the spectral profiles of $\rho_{vv}$ and $\rho_{ee}$, which is in turn directly proportional to the fluorescence signal, provides a convenient metric for the occurrence of a Raman transition, which we exploit in our experiment.

\begin{figure}[t]
\centering 
 \includegraphics[width=0.85\columnwidth]{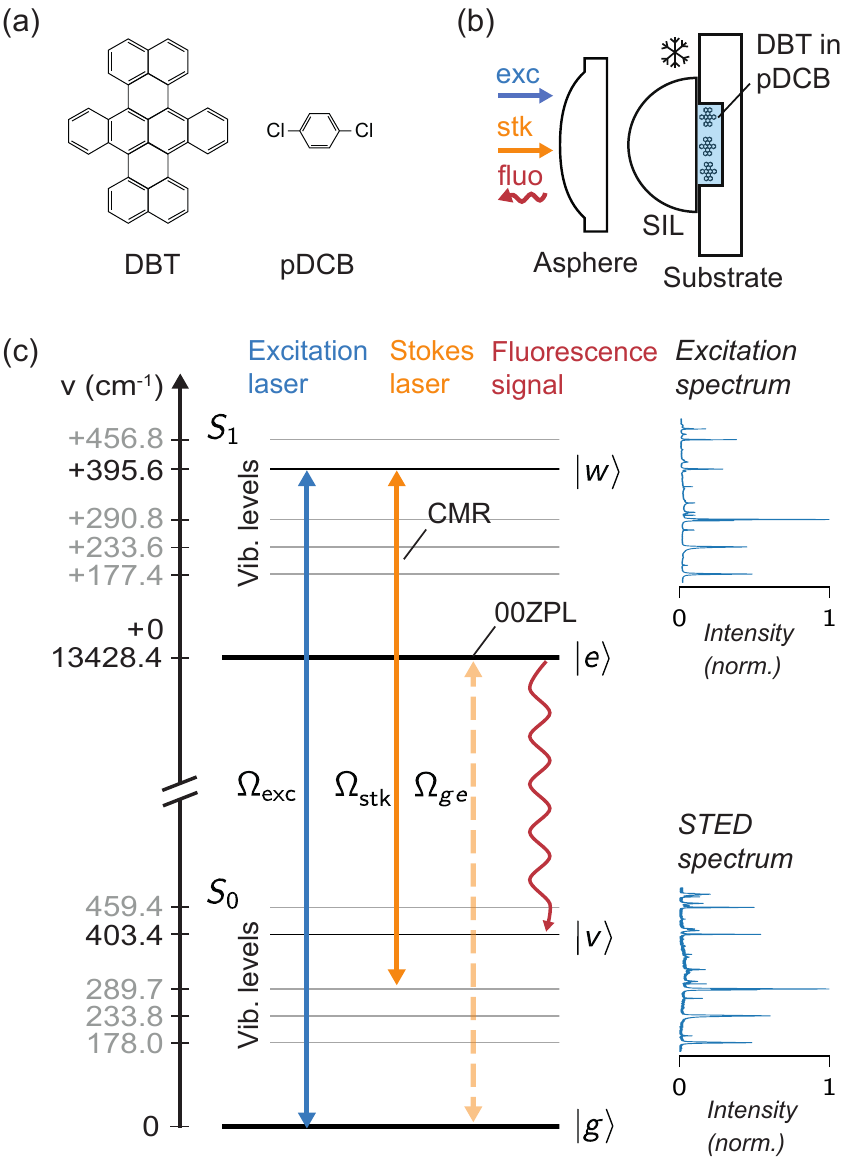}
 \caption{(a) Chemical structures of DBT and pDCB.
 (b) Sketch of the sample and focusing optics in the cryostat.
Exc: excitation laser, Stk: Stokes laser, Fluo: fluorescence, SIL: solid-immersion lens.
 (c) Level scheme of a DBT molecule with several vibronic states associated with the most prominent vibronic transitions from the vibrational ground states $\vert g \rangle$ and $\vert e \rangle$.
Transitions $\vert g \rangle \leftrightarrow \vert w \rangle$ and $\vert v \rangle \leftrightarrow \vert w \rangle$ addressed by the two lasers are marked by solid double-headed arrows. CMR: common mode resonance.}
\label{levels}
\end{figure}

The measurements in this work were performed on dibenzoterrylene (DBT: C$_{38}$H$_{20}$) embedded in a para-dichlorobenzene (pDCB: C$_6$H$_4$Cl$_2$) crystal \cite{verhart-2016} at a molar concentration of \SI{30}{ppb} (see Fig.\,\ref{levels}(a)). 
As sketched in Fig\,\ref{levels}(b), the doped pDCB crystal was formed in a channel of sub-micrometer depth created between a fused silica substrate and a solid-immersion lens (refractive index $n=2.14$) \cite{maser-2016, gmeiner-2016}. This configuration allowed efficient coupling of two focused laser beams to DBT molecules in the sample. We remark that we chose the DBT doping level to be low enough to find no more than one molecule in the excitation volume. Second-order autocorrelation measurements of the molecular fluorescence confirmed that the signals presented in this study stem from a single molecule  (see \cite{zirkelbach-2022}).
Experiments were performed in a closed-cycle dilution refrigerator that operated at a base temperature of $T\sim\SI{25}{mK}$ (see SM, section I). 

In Fig.\,\ref{levels}(c), we display the energy levels of the most prominent vibronic transitions associated with the electronic ground ($S_0$) and excited ($S_1$) states of a molecule according to the experimentally recorded data presented on the right-hand side of the figure \cite{zirkelbach-2022}. The linewidth of the 00ZPL connecting the electronic ground state $\vert g \rangle$ to the electronic excited state $\vert e \rangle$ with no vibrational excitation corresponds to $\Gamma_{e} /(2\pi) = \SI{23}{MHz}$. The linewidths of $\vert v \rangle$ and $\vert w \rangle$ were measured to be $\Gamma_{v} / (2\pi) = \SI{13.5}{GHz}$ and $\Gamma_{w} / (2\pi) = \SI{26.1}{GHz}$ by using incoherent spectroscopy schemes, namely, fluorescence excitation and stimulated emission depletion (STED), respectively (see Ref.\,\cite{zirkelbach-2022} and SM, section V). In order to drive a coherent Raman process, the beams of two independently tunable and narrowband (linewidth $<\SI{100}{kHz}$) Ti:Sapphire lasers were overlapped and coupled to a confocal microscope (see SM, section I). 

To optimize the efficiency of coherent transitions, it is desirable to work with high FC factors and low $\Gamma_{i}$. In molecules with a prominent 00ZPL (such as DBT), common mode resonances (CMR: electronic transitions conserving the vibrational quantum numbers \cite{carlson-1990}) tend to have the highest FC factors (see SM, section II). Moreover, it is favorable to choose a pair of vibronic states in $S_0$ and $S_1$ with a transition frequency that is considerably detuned from the 00ZPL to satisfy $\Omega_{ge} \ll \Delta_{ge}$ in order to minimize the probability of exciting $\vert e \rangle$ by the Stokes laser. We selected the CMR between the modes close to \SI{400}{cm^{-1}} with a detuning of about $\SI{232}{GHz}$ from 00ZPL. The response of the molecule was monitored via the red-shifted fluorescence on a prominent transition from $\vert e \rangle$ to a vibronic level in $S_0$ at about \SI{290}{cm^{-1}} (width of bandpass filter: $\SI{2}{nm}$). 

\begin{figure}[t]
\centering 
 \includegraphics[width=0.9\columnwidth]{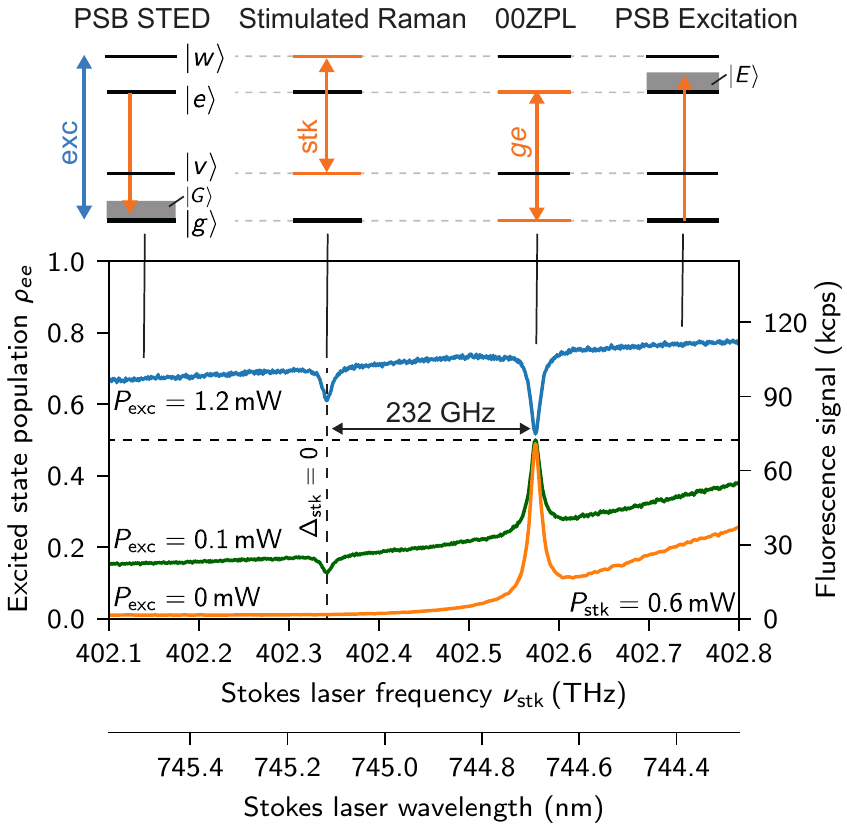}
 \caption{Scan of the Stokes laser frequency $\nu_{\rm stk}$ at various power levels of the resonant excitation laser ($\Dexc = 0$).
The level schemes on top of the plot indicate the processes underlying the spectral features in the data.
PSB: phonon sideband.
}
\label{stokes_scan}
\end{figure}

The spectra in Fig.\,\ref{stokes_scan} show the background-corrected fluorescence signal $R_{\mathrm{fluo}}$ and $\rho_{ee} = R_{\mathrm{fluo}} / R_{\infty}$ deduced from it as $\nu_{\rm stk}$ was scanned, while $\nu_{\rm exc}$ was set on the resonance of the $\vert g \rangle \leftrightarrow \vert w \rangle$ transition ($R_{\infty}$ denotes the fluorescence count rate at saturation; see SM, section IV). The scan range was chosen to cover both 00ZPL and $\vert v \rangle \leftrightarrow \vert w \rangle$ transition frequencies. Three exemplary spectra taken at different excitation powers represent the cases of negative and positive population inversion between $\vert g \rangle$ and $\vert e \rangle$. The level schemes in the upper part of the figure illustrate the different transitions that contribute to the various parts of the spectrum. In the case of $P_{\rm exc}=0$ (orange), we observe a peak when the Stokes laser is resonant with 00ZPL. The slope of the base line in the region where $\nu_{\rm stk} > \nu_{ge}$ is caused by excitation to the tail of the phonon side band (PSB) in $\vert e \rangle$ (denoted by $\vert E \rangle$) \cite{rebane-1993, clear-2020}. As the excitation power is increased (green), more population is transferred from $\vert g \rangle$ to $\vert e \rangle$ via $\vert w \rangle$ and a larger fluorescence signal is generated, leading to a higher base line. On the Raman   resonance ($\Dstk = \Dexc = 0$), the coherence induced between $\vert g \rangle$ and $\vert v \rangle$ leads to a dip in the fluorescence profile and the transfer of population to $\vert v \rangle$. At very large excitation powers ($\Pexc = \SI{1.2}{mW}$, blue), $\rho_{ee}$ exceeds $\rho_{gg}$ such that the Stokes laser can stimulate transitions back to $\vert g \rangle$, resulting in a dip at the 00ZPL frequency. At $\Pexc = \SI{0.1}{mW}$ and $\Pexc = \SI{1.2}{mW}$, the (slight) drop in the fluorescence signal for $\nu_{\mathrm{stk}} \lesssimsmall \SI{402.4}{THz}$ is related to stimulated emission from $\vert e \rangle$ to the PSB of $\vert g \rangle$ (denoted $\vert G \rangle$). 

\begin{figure}
\centering 
 \includegraphics[width=.95\columnwidth]{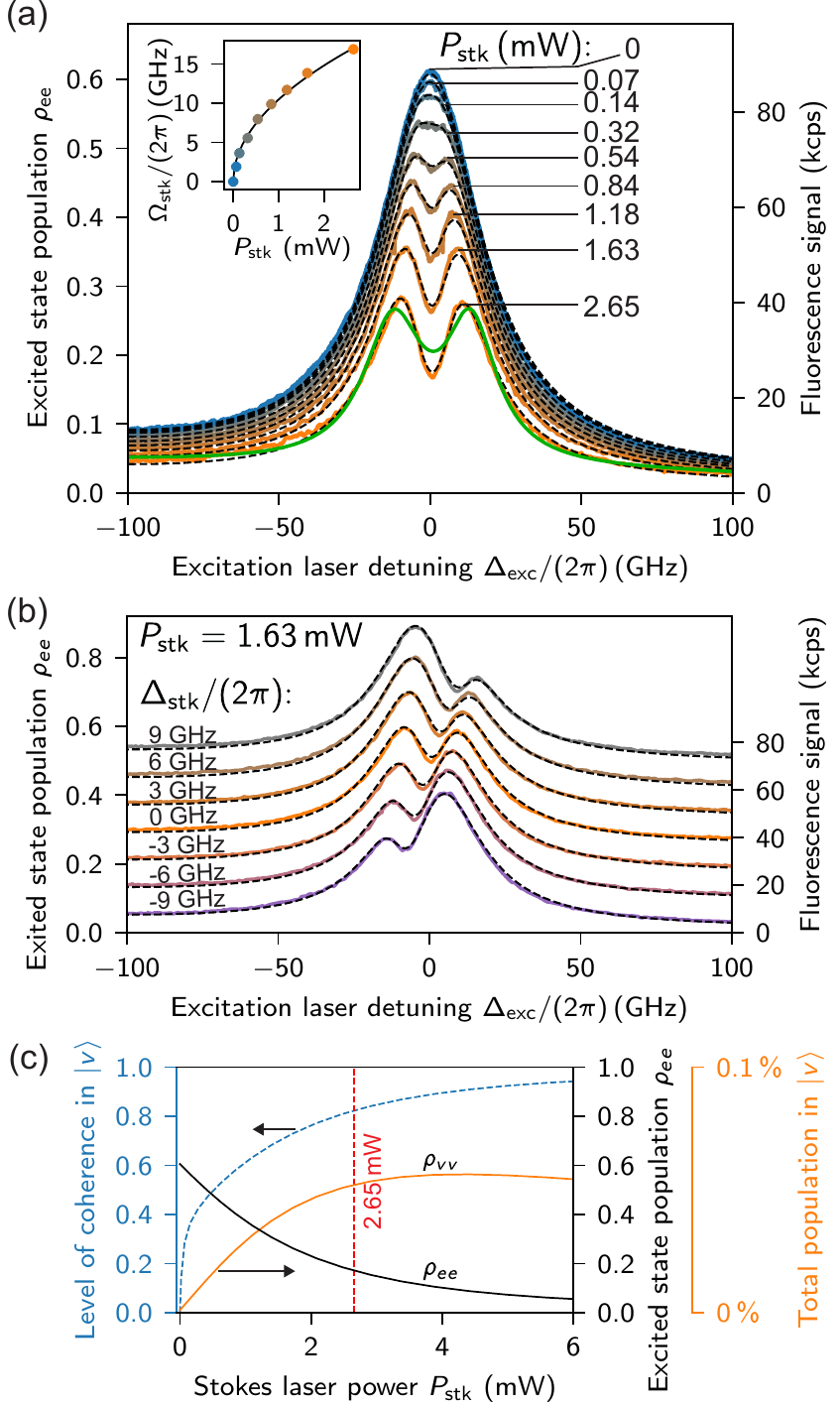}
 \caption{Resonant stimulated Raman transition in a single DBT molecule.
(a) Excitation spectra of the vibronic transition $\vert g \rangle \leftrightarrow \vert w \rangle$ monitored via fluorescence from $\vert e \rangle$ as a function of the excitation laser frequency detuning ($\Dexc$). Spectra were recorded at different powers ($\Pstk$) of the Stokes laser at $\Dstk = 0$ and $\Pexc = \SI{0.6}{mW}$ ($\Oexc = \sqrt{1.5 \Gamma_{e} \Gamma_{w}}$). Green curve depicts a fit to two Lorentzian functions.
(b) Same as in (a) but for different $\Dstk$ and at $\Pstk = \SI{1.63}{mW}$.
The data are plotted with an offset.
(c) Total populations $\rho_{vv}$ (solid orange line) and $\rho_{ee}$ (solid black line), 
and level of coherence in $\vert v \rangle$ ($\vert \rho_{vg}\vert^2 / \rho_{vv}$, dashed blue line) at $\Dexc = 0$ as predicted by the model fitted to the data in (a).} \label{splitting}
\end{figure}

Next, we fixed the frequency of the Stokes laser to the CMR ($\Dstk = 0$), set $\Pexc = \SI{0.6}{mW}$ (corresponding to $\Oexc \approx \sqrt{1.5 \Gamma_{e} \Gamma_{w}}$) and scanned $\nu_{\mathrm{exc}}$ through the resonance of the $\vert g \rangle \leftrightarrow \vert w \rangle$ transition. The outcome of these measurements is presented for various $\Pstk$ in Fig.\,\ref{splitting}(a) (see section IV of SM for background correction applied to the data). It is evident that as predicted in Fig.~\ref{model}(b), the original Lorentzian line profile develops a dip for larger $P_{\rm stk}$ values. This dip corresponds to
the inverse band shapes encountered in conventional resonant stimulated Raman spectroscopy \cite{saikan-1985, takayanagi-1988, pei-2013, wei-2018, shi-2018}. 

To compare our experimental data with theoretical expectations, the model presented in Fig.\,\ref{model} was modified to include energy transfer to the crystal via $\vert G\rangle$ and $\vert E \rangle$ as well as an additional vibronic transition to a level $\vert x \rangle$ close to $\vert w \rangle$ (see SM, section III). The black lines in Fig.\,\ref{splitting}(a) show the results of the model fit to the data, whereby only $\Ostk$ and $\beta_{G} =  \lVert \vec{d}_{Ge} \rVert / \lVert \vec{d}_{vw} \rVert$ acted as free parameters. All remaining quantities 
were independently determined from STED spectroscopy \cite{zirkelbach-2022} and a fluorescence excitation scan in the absence of the Stokes laser (see SM, section V).  
As expected, we find that $\Ostk$ scales with $\sqrt{\Pstk}$ (see inset in Fig.\,\ref{splitting}(a)).  The drops in the fluorescence amplitude and baseline are caused by stimulated depletion of the excited state into $\vert G \rangle$ (see  sections III and V in SM). We remark the observed dip cannot be attributed to an ATS process as the spectral profile clearly deviates from a fit to the sum of two Lorentzians (see the green curve in Fig.\,\ref{splitting}(a)) \cite{anisimov-2011,lu-2015, hao-2018}. 

We also performed a similar experiment while varying the Stokes frequency detunings around the CMR at a fixed Stokes laser power ($\Pstk = \SI{1.63}{mW}$). The outcome is presented in Fig.\,\ref{splitting}(b) and reports a smooth shift of the induced dip over the spectral profile of the $\vert g \rangle \leftrightarrow \vert w \rangle$ transition. The dashed curves display the result of the theoretical model, which is in excellent agreement with the measured data without any additional free fit parameters. 

In Fig.\,\ref{splitting}(c), we plot the predictions of the model for populations $\rho_{vv}$ (orange curve) and $\rho_{ee}$ (black curve) when $\Dexc = \Dstk = 0$. The blue dashed line presents the degree of coherence $\vert \rho_{gv} \vert^2 / \rho_{vv}$ obtained from the fit to the data of Fig.\,\ref{splitting}(a). The model states that for the maximum power $\Pstk = \SI{2.65}{mW}$ used in our experiment (red dashed line), approximately 80\% of the population that is transferred to $\vert v \rangle$ undergoes a coherent process (see SM, section IV). In other words, the molecular vibrations have a fixed phase relation with the beating oscillations of the two driving lasers. This phenomenon is analogous to the coherence of scattering from a laser-driven two-level atom \cite{tannoudji-1992}. We also find that the level of coherence transferred to $\vert v \rangle$ increases with $\Pstk$ as the laser-mediated transfer becomes faster than $\Gamma_v$ and $\Gamma_w$. Moreover, $\rho_{vv}$ drops again with increasing $\Pstk$ due to the emergence of a line splitting.  Since $\vert v \rangle$ relaxes within $\sim \SI{10}{ps}$, the steady-state population in this level remains limited to $\rho_{vv} < 10^{-3}$ in the experimental scheme employed here. 

\begin{figure}
\centering 
 \includegraphics[width=.9\columnwidth]{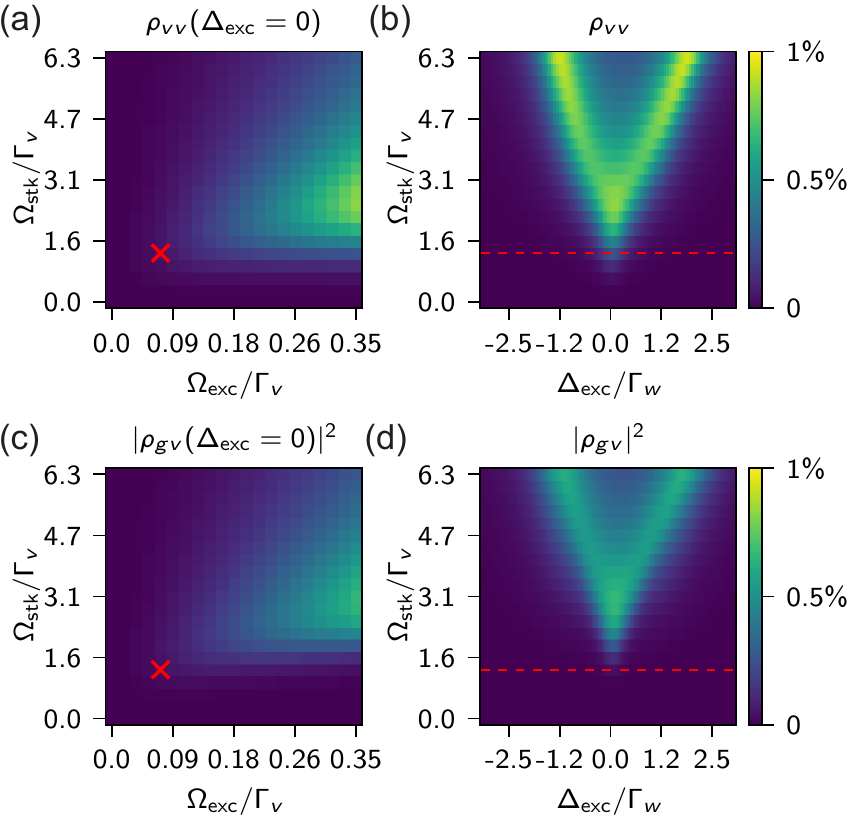}
 \caption{Theoretical predictions for total (a,b) and coherent (c,d) population in $\vert v \rangle$ for various combinations of the excitation and Stokes laser Rabi frequencies ($\Oexc$ and $\Ostk$) as well as the excitation laser frequency detuning. 
 $\Dstk = 0$ for all cases. 
 The red markers and lines indicate the (maximum) values used in our experiments.
 } 
 \label{predictions}
\end{figure}

To elucidate the wider parameter space that can be explored, we present the vibrational population $\rho_{vv}$ and its coherent part $\vert\rho_{gv}\vert^2$ as a function of $\Omega_{\mathrm{exc}}$, $\Omega_{\mathrm{stk}}$, and $\Delta_{\mathrm{exc}}$ in Fig.\,\ref{predictions}. The red crosses and dashed lines mark the parameters of this work. The data in Fig.\,\ref{predictions}(a,c) show that on resonance, a considerably more efficient population transfer could take place if one increased the excitation power by about 25 times. In Fig.\,\ref{predictions}(b,d), we study the effect of frequency detuning away from resonance, emphasizing that due to the splitting caused by the coherent interaction, the optimal Raman transition does not take place at the original resonance condition $\Dexc = 0$. Furthermore, Fig.\,\ref{predictions}(b) shows the regime, where a population transfer as large as 1\% to state $\vert v \rangle$ can be achieved (of which 60\% is coherent), corresponding to an efficient vibrational Raman process in a molecular system. 

In conclusion, we studied the laser-induced coherent line splitting associated with a stimulated Raman transition in a single organic molecule embedded in a crystal. Background fluorescence was minimized by probing a higher-lying vibrational level that does not directly fluoresce at the detection wavelength. Furthermore, the Stokes laser was tuned to a common mode resonance with a high FC factor to compete with the fast relaxation rates of the vibronic states in the solid state. By using tight focusing at cryogenic temperatures, the Raman transition could be achieved at CW laser powers as low as \SI{1}{mW}. 

Our study paves the way for a number of future investigations. For example, the observed line splitting can be used to infer the absolute transition dipole moments of vibronic transitions if a thorough intensity calibration of the optical focus is performed \cite{qi-2002, ahmed-2006}. Furthermore, one can study the Raman effect in single molecules in a quantitative fashion and in close comparison with theoretical models, e.g., as one tunes the transition from resonance to far off resonance conditions \cite{mukamel-1995, shi-2018, wei-2018}. Moreover, the rich manifolds of electronic and vibrational energy levels in molecules \cite{tesch-2002, suzuki-2005, roelli-2016, bayliss-2020, gurlek-2021} could be explored for achieving coherent population transfer and storage for applications in quantum information processing, similar to schemes such as STIRAP \cite{bergmann-2015} or EIT \cite{fleischhauer-2005, phillips-2001, hoeckel-2010, mucke-2010, specht-2011, slodivcka-2010}. 


\bibliography{zirkelbach-lib} 

\end{document}